\documentclass[12pt]{iopart}

\def\Journal#1#2#3#4{{#4} {\it #1} {\bf #2}, #3 }

\begin{document}

\letter{Purely gravito-magnetic vacuum space-times}

\author{Norbert Van den Bergh}

\address{Faculty of Applied Sciences TW16, Gent University, Galglaan 2, 9000 Gent, Belgium}

\begin{abstract}
It is shown that there are no vacuum space-times (with or without cosmological constant) for which the
Weyl-tensor is purely gravito-magnetic with respect to a
normal and timelike congruence of observers.

\end{abstract}

\pacs{0420}



\section{Introduction}
Non-conformally flat space-times for which the metric is an exact
solution of the Einstein field equations
\begin{equation}
G_{ab}\equiv R_{ab}-\frac{1}{2}R g_{ab}+\Lambda g_{ab}=T_{ab}
\end{equation}
and in which there exists a family of observers with 4-velocity
$u^a$ ($u_au^a=-1$) such that the gravito-electric part of the
Weyl-tensor vanishes,
\begin{equation}\label{E_ab}
E_{ab}\equiv C_{abcd} u^b u^d=0,
\end{equation}
are called purely gravito-magnetic space-times. The remaining
gravito-magnetic part of the Weyl-tensor,
\begin{equation}
H_{ab}\equiv C^*_{abcd}u^b u^d ,
\end{equation}
has no Newtonian analogue and its role in the dynamics of the
gravitational field is not very clear, apart from the fact that it
is associated with gravitational radiation~\cite{Maartens1,Hogan,
Dunsby}. Although purely gravito-magnetic space-times are the
subject of some intensive current
research~\cite{Trumper,Arianrhod,McIntosh,Haddow,Maartens,Lozanovski,Ferrando},
only a handfull of models with a reasonable matter source is
known, a situation which is in stark contrast to the purely
gravito-electric space-times, for which wide and physically
important classes of examples exist. This is particularly true for
the vacuum solutions $T_{ab}=0$, where for example all the static
vacua are purely gravito-electric, while no purely
gravito-magnetic solutions are known at all! This has lead some
researchers to conjecture that purely gravito-magnetic vacua do
not exist~\cite{McIntosh, Maartens}, but this so far has only been
proved in the special cases where the Petrov type is
D~\cite{McIntosh} or where the timelike congruence $u^a$ is normal
and shear-free~\cite{Trumper}. In the present letter it is shown
that the conjecture is true provided that the congruence is normal
only .

\section{Dynamical equations}

For a vacuum purely gravito-magnetic space-time in which the
timelike congruence $u^a$ is normal, one of the constraint
equations (the "divergence of $\mathbf{E}$" equation, see for
example \cite{Maartens}) guarantees that the shear tensor
$\sigma_{ab}$ commutes with $H_{ab}$ (note that it makes sense to
talk about \emph{the} kinematical quantities, as these space-times
are necessarily of Petrov-type I~\cite{McIntosh} and, as can be
easily seen, the timelike congruence is then uniquely defined).
As both tensors are
orthogonal to the timelike congruence it follows that an
orthonormal tetrad (with $\mathbf{u}=\mathbf{e}^0)$ exists in
which $\sigma_{ab}$ and $H_{ab}$ are diagonal. From now on I will
follow the  notations and conventions of the orthonormal tetrad
formalism~\cite{MacCallum}, with the exception of the coefficients
$n_{\alpha \alpha}$ being redefined as follows:
\begin{equation}
n_{11}=(n_2+n_3)/2,\ n_{22}=(n_3+n_1)/2, \ n_{33}=(n_1+n_2)/2
\end{equation}
As the system of equations is SO(3)-invariant, each triplet of
equations will be represented by a single equation (the others
being obtained by cyclic permutation of the indices). The
vanishing of the gravito-electric part of the Weyl-tensor can then
be expressed by the 9 equations
\begin{eqnarray}
\fl E_{11}\equiv -\partial_0 \theta_1-\theta_1^2+\partial_1
{\dot u}_1 + {\dot u}_1 ^2- {\dot u}_2 (a_2-n_{13})- {\dot u}_3
(a_3+n_{12}) +\frac{1}{3} \Lambda =0 \\
\fl E_{12}\equiv \partial_2 {\dot u}_1 + {\dot u}_2 ( {\dot u}_1
+n_{23}+a_1)+\frac{1}{2} {\dot u}_3 n_2 +\Omega_3 (\theta_2-\theta_1) =0 \\
\fl E_{21}\equiv \partial_1 {\dot u}_2 + {\dot u}_1 ( {\dot u}_2 -
n_{13}+a_2)-\frac{1}{2} {\dot u}_3 n_1 +\Omega_3 (\theta_2-\theta_1) =0
\end{eqnarray}
The vanishing of the off-diagonal components of $H_{ab}$ on the other hand leads to
\begin{eqnarray}
\fl H_{12}\equiv -\partial_0(n_{12}+a_3) - \partial_1 \Omega_2 -
\theta_1 (n_{12}+a_3+ {\dot u}_3 )\nonumber \\
+\Omega_1 (n_{13}-a_2) -\Omega_2 (n_{23}-a_1+ {\dot u}_1 )
+\frac{1}{2} \Omega_3 (n_1-n_2) =0 \\
\fl H_{21}\equiv -\partial_0(n_{12}-a_3) - \partial_2 \Omega_1 -\theta_2 (n_{12}-a_3- {\dot u}_3 )\nonumber \\
+\Omega_1 (n_{13}+a_2-{\dot u}_2) -\Omega_2 (n_{23}+a_1 )
+\frac{1}{2} \Omega_3 (n_1-n_2) =0
\end{eqnarray}
It might strike as odd that the off-diagonal components of
$E_{ab}$ and $H_{ab}$ are listed as independent equations, as both
are symmetric tensors. This is the price one has to pay (or
perhaps the benefit which is obtained?) by steering away from the
covariant approach: the symmetry of $E_{ab}$ and $H_{ab}$ is now
guaranteed by the Jacobi-identities. This becomes clear when we
use the previous expressions to obtain (a) evolution equations for
$\theta_\alpha$, $a_\alpha$, $n_{\alpha \beta}$ and (b)
expressions for the spatial gradients $\partial_\alpha {\dot
u}_\beta$ ($\alpha \not = \beta$) of the acceleration. Of the 16
Jacobi-identities and the 6 $(0 \alpha)$ and $(\alpha \beta)$
($\alpha \not = \beta$) components of the field equations only 15
algebraically independent equations remain. These equations can be
written as evolution equations for the coefficients $n_\alpha$,
\begin{equation}
\fl \partial_0 n_1 = 2 \partial_1 \Omega_1 + 2 {\dot u}_1 \Omega_1
+4 (\Omega_2 n_{13}- \Omega_3 n_{12}) -n_1
\theta_1-n_2(\theta_1-\theta_3)-n_3 (\theta_1-\theta_2)
\end{equation}
and as expressions for the spatial gradients of $\theta_\alpha$
and $n_\alpha$:
\begin{eqnarray}
\partial_1 \theta_2 = (\theta_2-\theta_1) (n_{23}+a_1) \\
\partial_2 \theta_1 = (\theta_2-\theta_1) (n_{13}-a_2) \\
\partial_1 n_2 = -2 \partial_2(n_{12}+a_3)-4 n_{12} (n_{13}-a_2)-2 n_{23} n_1+2 a_1 n_2 \\
\partial_2 n_1 = -2 \partial_1(n_{12}-a_3)+4 n_{12} (n_{23}+a_1)+2 n_{13} n_2+2 a_2 n_1
\end{eqnarray}
Finally, the relation between the diagonal components of $H_{ab}$
and the rotation coefficients is given by
\begin{equation}\label{mag}
2 H_{11}=n_2 \theta_3+n_3 \theta_2 -(n_2+n_3) \theta_1
\end{equation}

\section{Propagating the Einstein equations}
The equations of the previous section allow us now to propagate
the diagonal components of the Einstein equations along
$\mathbf{u}$. While the (00) component is an identity, the
$(\alpha \alpha )$ components yield
\begin{eqnarray}\label{R11}
\fl 2 \partial_1 a_1 +\partial_2 a_2 +\partial_3 a_3+\partial_3 n_{12}-
\partial_2 n_{13} \nonumber \\
= 2(n_{23}^2+a_1^2+a_2^2+a_3^2)-2(a_2 n_{13}+a_3 n_{12})
-\frac{1}{2}
n_2 n_3 \nonumber \\
- \theta_1 (\theta_2+\theta_3)+\frac{2}{3}
\Lambda
\end{eqnarray}
The trace of these  equations is given by
\begin{eqnarray}\label{trace}
\fl 4(\partial_1 a_1 +\partial_2 a_2 +\partial_3 a_3) = 2
(n_{12}^2+n_{23}^2+n_{13}^2)
+6 (a_1^2+a_2^2+a_3^2) \nonumber \\
-\frac{1}{2} (n_1n_2+n_2n_3+n_3n_1)-2(\theta_1\theta_2+\theta_2
\theta_3 +\theta_3 \theta_1)
\end{eqnarray}
Acting now with the $\partial_0$ operator on (\ref{R11}) one can
eliminate the second order derivatives of the coefficients
$a_\alpha$ by using the evolution equations of the previous
section together with the commutator relations $[\partial_0 ,\
\partial_\alpha ] a_\alpha$. What complicates matters at this stage is the
introduction of various second order \emph{spatial} derivatives of
the $\Omega_\alpha$ (which describes the rotation of the spatial
triad with respect to a Fermi-propagated frame). A small miracle
however ensures (a) that these derivatives always appear under the
form of $[\partial_\alpha ,\
\partial_\beta ] \Omega_\gamma$ commutators and hence can be
eliminated and (b) that the remaining first order derivatives of
$\Omega_\alpha$ all cancel out. Furthermore the resulting equation
contains the $a_\alpha$ derivatives only in the "divergence of
$\mathbf{a}$" form, which is known from (\ref{trace}). The final
result is a remarkably simple triplet of equations:
\begin{equation}\label{eq1}
\fl (n_2-n_3)n_1 (\theta_3-\theta_2) - n_2n_3 (2 \theta_1-\theta_2-\theta_3) -2 n_2^2 (\theta_1-\theta_3) - 2 n_3^2 (\theta_1-\theta_2)=0
\end{equation}
Provided that the shear-tensor is not degenerate (if this would be
the case, then substituting e.g. $H_2=H_1$ in (\ref{eq1}) and its
cyclic permutations shows that also $H_{ab}$ is degenerate, such
that the solution would be of Petrov type D), one can eliminate
the coefficients $n_\alpha$ from equations (\ref{eq1}) and
(\ref{mag}) and their cyclic permutations, to obtain
\begin{equation}
H_{11}^2+H_{11}H_{22}+H_{22}^2=0
\end{equation}
which is clearly inconsistent with the fact that the Petrov type is I.


\section*{References}
\begin{thebibliography}{99}
\bibitem{Maartens1} Maartens R, Ellis G F R and Siklos S T \Journal{Class. Quantum Grav.}{14}{1927}{1997}
\bibitem{Hogan} Hogan P and Ellis G F R \Journal{Class. Quantum Grav.}{14}{A171}{1997}
\bibitem{Dunsby} Dunsby P K S, Bassett B A and Ellis G F R \Journal{Class. Quantum Grav.}{14}{1215}{1997}
\bibitem{Trumper} Tr\"umper M \Journal{J. Math. Phys.}{6}{584}{1965}
\bibitem{Arianrhod} Arianrhod R, Lun A W-C, McIntosh C B G and Perj\'es Z
\Journal{Class. Quantum Grav.}{11}{2331}{1994}
\bibitem{McIntosh} McIntosh C B G, Arianrhod R, Wade ST and Hoenselaers C
\Journal{Class. Quantum Grav.}{11}{1555}{1994}
\bibitem{Haddow} Haddow B M {\it Purely Magnetic Spacetimes}\ \  gr-qc/9502041
\bibitem{Maartens} Maartens R, Lesame W M and Ellis G F R
\Journal{Class. Quantum Grav.}{15}{1005}{1998}
\bibitem{Lozanovski} Lozanovski C and Aarons M
\Journal{Class. Quantum Grav.}{16}{4075}{1999}
\bibitem{Ferrando} Ferrando J J and S\'aez J A
\Journal{Class. Quantum Grav.}{19}{2437}{2002}
\bibitem{MacCallum} MacCallum M A H 1971 {\it Cosmological Models from a Geometric Point of View
(Carg\`ese)}~Vol 6 (New York: Gordon and Breach) p 61


\end{thebibliography}
\end{document}